# CAUSE AND EXTENT OF THE EXTREME RADIO FLUX DENSITY REACHED BY THE SOLAR FLARE OF 2006 DECEMBER 06

Dale E. Gary

Center for Solar-Terrestrial Research, Physics Department, New Jersey Institute of Technology,
323 M. L. King Jr. Blvd., Newark, NJ 07102

**Abstract.** The solar burst of 2006 December 06 reached a radio flux density of more than 1 million solar flux units (1 sfu = $10^{-22}$ W/m$^2$/Hz), as much as 10 times the previous record, and caused widespread loss of satellite tracking by GPS receivers. The event was well observed by NJIT's Owens Valley Solar Array (OVSA). This work concentrates on an accurate determination of the flux density (made difficult due to the receiver systems being driving into non-linearity), and discuss the physical conditions on the Sun that gave rise to this unusual event. At least two other radio outbursts occurred in the same region (on 2006 December 13 and 14) that had significant, but smaller effects on GPS. We discuss the differences among these three events, and consider the implications of these events for the upcoming solar cycle.

**1. Introduction**

During the period 2006 December 5-14, a period near the minimum of the solar activity cycle, active region (AR) 10930 produced a remarkable series of X-class flares. Although the X-ray flux of the flares was not exceptional, several of the events produced record radio noise in the L-band (1-2 GHz frequency range). What makes these events of special interest are the reports of widespread effects on Global Positioning System (GPS) receivers. The potential for such effects caused by solar radio noise during flares was discussed by Klobuchar (1999) and Chen (2005), and the direct quantitative comparison that demonstrated cause and effect was shown by Cerruti et al. (2006) for a burst in 2003. As a result of these studies, it was thought that solar radio bursts could have small but non-negligible effect on GPS receivers, but the situation changed on 2006 December 06, when a soft-X-ray class X6.5 event caused widespread failure of GPS receiver position determination over the entire sunlit hemisphere of Earth (Cerruti et al. 2008). The extent of the GPS effects and the failure mode of the GPS receivers is detailed elsewhere for several of the 2006 December bursts (Cerruti et al. 2008; Afraimovich et al. 2007, 2008; Kintner 2008). This paper focuses on the question of how the Sun produced such strong radio emission in the L-band frequency range, and how often such events can be expected to occur in the future.

To answer the question of how often such high-flux events will occur in the future, it is important to know the historical record of past radio bursts. Solar radio emission is monitored by a world-wide network of radiometers, whose measurements are reported by the National Oceanic and Atmospheric Administration (NOAA) National Geophysical Data Center (NGDC). Nita et al. (2002) reported on the spectral and solar-cycle dependence of the bursts in this data set, extending over 40 years from 1960-2000. This is an inhomogeneous data set made by dissimilar instruments, although in later years the reports are dominated by the relatively homogeneous US Air Force Radio Solar Telescope Network (RSTN). The RSTN instruments have stated flux density limits of $5\times10^5$ sfu (solar flux units; 1 sfu = $10^{-22}$ W m$^{-2}$ Hz$^{-1}$) below 1 GHz, but only $10^5$ sfu at L band (1415 MHz), and $5\times10^4$ sfu above 2 GHz. As we will see, the reported RSTN flux densities are lower than the true ones for the 2006 Dec 6, 13 and 14 events. This suggests that the largest bursts may be missing from the historical record due to saturation of the measurements. Indeed, Nita et al. (2002) pointed out in their study of the size distribution of solar bursts that their results are consistent with an undercount of large events, although they could not say whether this was the result of instrument saturation or whether the Sun simply does not produce large events. With the bursts of 2006 December, we can now say with certainty that the Sun can indeed produce events that exceed the RSTN saturation level.

To put the 2006 December bursts in perspective, in §2 we give an overview of the flares produced by AR 10390, including the 4 X-class flares, and describe their associated radio bursts. We point out that the last three of these bursts, 2006 Dec. 6, 13 and 14, all produced exceptionally high L-band flux density, although their radio flux densities were not atypical at other frequencies, and that all three are associated with GPS effects. In §3, we examine the issue of saturation of the measurements, to show that the extreme flux density of these bursts challenged the ability of current instruments to make accurate flux density measurements. In §4 we discuss the physical cause of the high L-band flux density, especially for the well-observed 2006 December 6 event, and show that it is due to coherent plasma emission involving a type of burst called ms spike bursts (Slottje 1978), thought to be due to the Electron-Cyclotron Maser (ECM) mechanism (see Treumann 2006 and references therein). We conclude in §5 with a discussion of the implications of the 2006 December bursts for assessing threats to GPS and other navigation systems.

## 2. The 2006 December Flares and Associated Radio Bursts

AR 10390 was already formed and growing when it rotated over the East limb of the Sun on 2006 December 5. On this date, it produced the largest of the X-class soft X-ray flares, an X9.0 peaking at 10:35 UT. It produced a second X-class flare the next day, an X6.5 peaking around 18:45 UT on 2006 December 6. Two more X-flares occurred a week later, on December 13 and 14. We will refer to these four X-class flares as Flares 1-4, in chronological order of their occurrence. Figure 1 shows the history of the region's flare production during its 2-week rotation across the solar disk. Although the region was highly flare-productive at times, it was by no means a record flare producer, either in number of flares or their soft X-ray class.

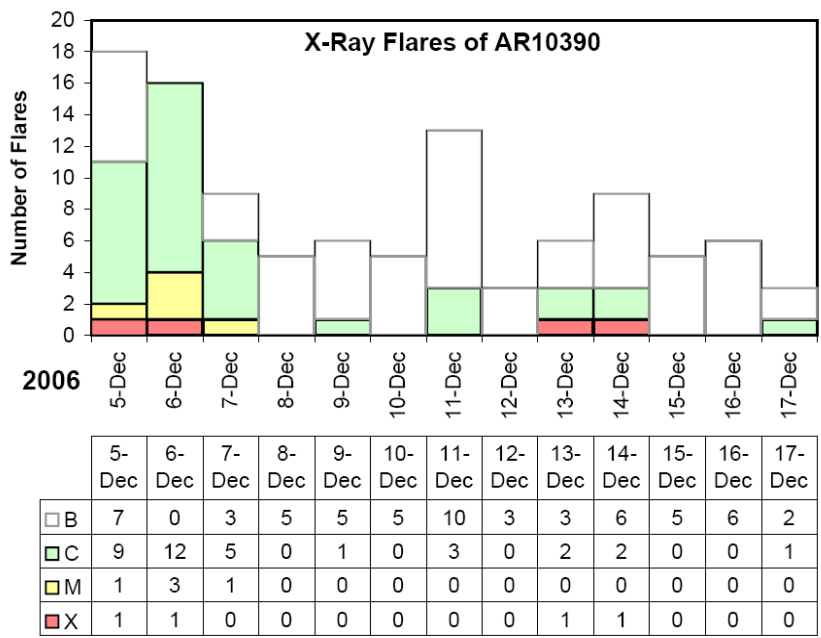

**Figure 1:** Histogram of occurrence of flares of various soft X-ray class, B (white), C (green), M (yellow) and X (red), versus date, starting on the day the region appeared on the East limb of the Sun and continuing for the approximately 2 weeks of passage across the solar disk. The region was quite flare productive the first few days, and then produced only a small number of C-class flares, with the major exception of two X-class flares on 13-14 December. B-class flares are typically too small to see during solar maximum, and would not even be reported were it not solar minimum.

In the production of radio noise, however, the region was a record breaker. The L-band flux density attained by both the 2006 December 6 and 13 events (~1 million and >300,000 sfu, respectively) exceeded the previous record (165,000 sfu for a burst in April 1973—see Cerruti et al. 2008) by large factors. The first event on 2006 December 5, in contrast, was not exceptional at L band despite the fact that it was the largest of the four X-class bursts as measured in soft X-rays.

The radio bursts were observed with several radio instruments. We report here on data from the four RSTN sites, the Owens Valley Solar Array (OVSA) and FASR Subsystem Testbed (FST), both in California, and the Nobeyama Radio Polarimeter (NoRP) in Japan. The RSTN sites each observed one of the flares, while OVSA and FST observed flares 2 and 4, and NoRP observed flare 3.

As an example that illustrates some of the issues, we first show the OVSA dynamic spectra for Flare 2 (Dec 6) in Figure 2. OVSA records data at 39 frequencies between 1.2 and 18 GHz, in both senses of circular polarization, with a time resolution of 8.1 s (see Nita et al. 2004). It is significant that GPS transmissions are right circularly polarized (RCP). One of the clear characteristics of the December bursts is that their L-band emission was also highly RCP. Solar bursts may be highly polarized in either sense of circular polarization (depending on the polarity of magnetic field in the source region), or may have little or no polarization. Only the RCP component of the solar emission can affect GPS receivers. Figure 2a shows radio flux density in a logarithmic scale, with black representing low radio flux density and white representing high flux density. The scale ranges from 1 to $10^4$ sfu, so emission greater than $10^4$ sfu is saturated. Note especially the strong, saturated emission below 2 GHz in the RCP, while in LCP it is unsaturated. The radio burst lasted about 1 h 40 m and, except for the strong L-band RCP emission, the burst is quite typical of large flares. OVSA maintains a high dynamic range by inserting attenuation in the re-ceiver to keep the signal in range, based on the measured flux density in the previously measured sample 8.1 s earlier. However, when the fluctuations in flux density are too rapid, the system may have too little attenuation and will then saturate. Such points are flagged as bad data, and appear black in Fig. 2. This is especially noticeable at the lowest frequency, 1.2 GHz.

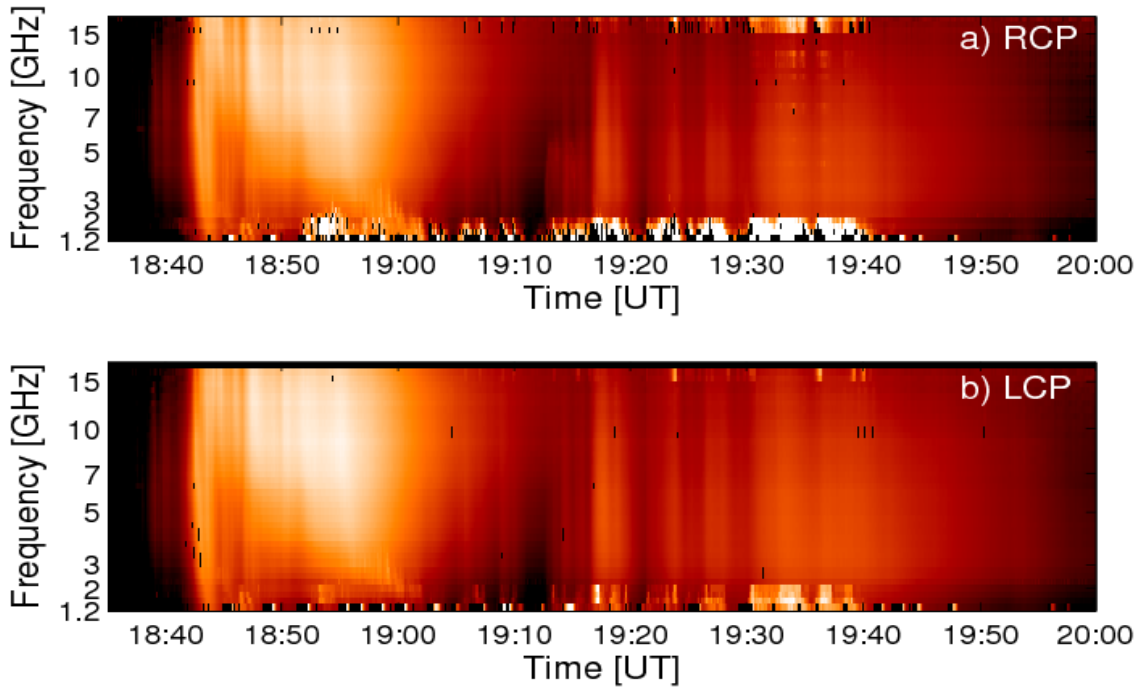

**Figure 2:** OVSA dynamic spectrum of the largest of the radio bursts, the 2006 Dec 06 event. *a)* right circular polarization (RCP). *b)* left circular polarization (LCP). The flux density scale ranges logarithmically from 1 to $10^4$ sfu, so the very bright RCP L-band emission (1-2 GHz) is saturated on this scale and appears white. True saturation (see text) causes the data to be flagged as bad data, and appears black in the figure, which is especially noticeable at 1.2 GHz. The brightest L-band RCP emission occurs between 19:30 and 19:40 UT, and causes instrumental artifacts at higher frequencies (between 15-18 GHz, and at times at other frequencies as well).

Fortunately, at the time of these flares a new instrument was operating at Owens Valley, the FASR Subsystem Testbed (Liu et al. 2007), which was observing at extremely high time and frequency resolution (20 ms and 977 kHz, respectively) in the band 1.0-1.5 GHz, so we can in effect zoom in on this portion of the spectrum as shown in Figure 3. The time resolution in Fig. 3 has been vastly reduced to match the OVSA time resolution, for display purposes, but we will use the full resolution in §4 where we discuss the physical origin of the emission. Although the emission in Fig. 3 appears relatively smooth and featureless, it in fact exhibits extreme variability at shorter time scales.

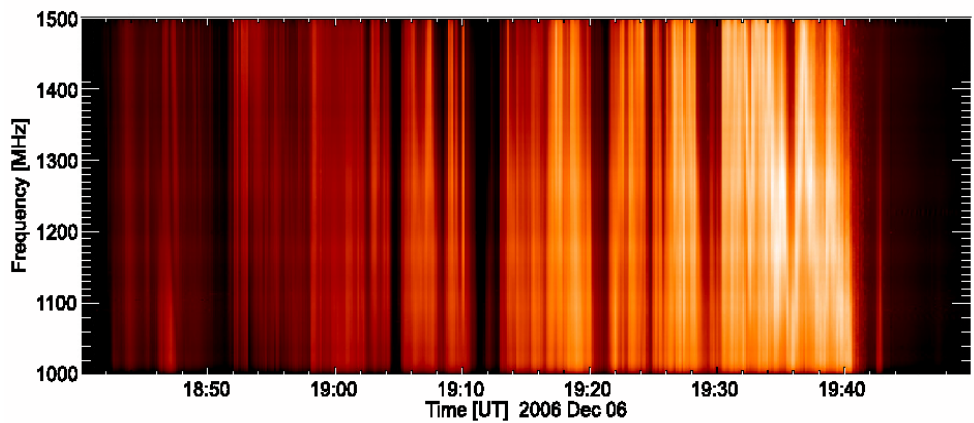

**Figure 3:** FST dynamic spectrum of the 1.0-1.5 GHz RCP emission, scaled logarithmically from 1 to $10^6$ sfu. The time resolution of the FST instrument has been vastly reduced, to 8 s to match the OVSA data, but it is actually far better, at 20 ms, allowing us to unambiguously determine the emission mechanism.

- 3 -

The main points to draw from Figs. 2 and 3 are that the burst emits over a broad range of frequencies, and that in the L-band it produced vast amounts of radio noise in RCP. GPS operates at two frequencies, the so-called L1 frequency at 1575 MHz, and L2 at 1227 MHz (see Cerruti et al. 2008 for more details), so this L-band emission is highly significant for GPS receivers.

The RSTN stations operate at 8 frequencies, 245, 410, 610, 1415, 2695, 4995, 8400, and 15400 MHz, and records data only in total intensity (no polarization information). The 1415 MHz frequency is the one of interest for potential impact on GPS and other navigation systems. Peak flux densities were reported for all four X-class events, from four different RSTN stations. As shown in Table 1, the RSTN flux was reported to be less than the peak L-band flux for three of the four events, but for different reasons that will be discussed in §3.

Table 1: Reported L-Band Flux vs. Actual Peak Flux for X-Class Events

| Flare Number | Date of X-Flare | Soft X-ray Class | RSTN Station | RSTN 1400 MHz Flux Reported | Actual L-band Peak Flux |
|---|---|---|---|---|---|
| **1** | 2006 Dec 05 | X9.0 | San Vito | 3,900 | 3,900 |
| **2** | 2006 Dec 06 | X6.5 | Sagamore Hill | 13,000 | ~1,000,000 |
| **3** | 2006 Dec 13 | X3.4 | Learmonth | 130,000 | 440,000 (1 GHz) |
| **4** | 2006 Dec 14 | X1.5 | Palehua | 2,700 | 120,000 |

The NoRP instrument operates at 7 frequencies, 1000, 2000, 3750, 9400, 17000, 35000 and 80000 MHz, and records both RCP and LCP flux densities. Only Flare 3 occurred during the NoRP observing time, and its L-band frequencies bracket the 1415 MHz RSTN frequency, so comparisons are inexact, but the reported flux densities of 440,000 sfu at 1 GHz and 302,000 sfu at 2 GHz suggest that the 1415 MHz flux density falls somewhere in the range $3\text{-}4\times10^5$ sfu.

RSTN radio flux density time profiles at 1415 and 2695 MHz are shown in Figure 4 for Flares 1 and 3, in order to contrast their radio behavior. Although the soft X-ray class of the X9.0 event on December 5 exceeds that of the X3.4 event on December 13, the order of their relative radio output is reversed. The smaller soft X-ray event (Flare 3) was a much greater producer of radio emission. We discuss this further in §4.

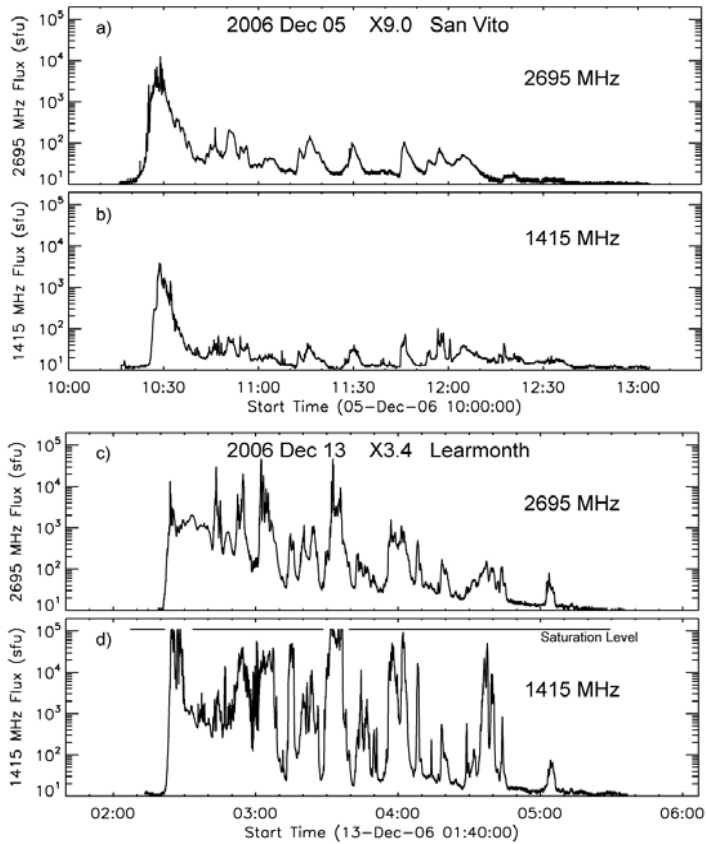

**Figure 4:** Radio flux density at two frequencies for Flares 1 and 3, recorded at RSTN sites. *a)* Flare 1, recorded on 2006 December 05 at San Vito, Italy, displays typical flux density levels for a large flare, reaching about 10,000 sfu at 2695 MHz, and showing a relatively smooth time profile. *b)* The same flare at 1415 MHz reached only 3900 sfu. *c)* Flare 3, recorded on 2006 December 13 at Learmonth, Australia, at 2695 GHz, displays rapid variations and stays at elevated flux levels for several hours. *d)* The same flare at 1415 MHz, showing that the flux density increases at lower frequencies, typical of the bursts affecting GPS systems. Note that the highest peaks at 1415 MHz are clipped at the 100,000 sfu saturation level of the RSTN receivers.

## 3. Effects of Saturation on Measurements of Radio Noise Level

In order to assess the threat of solar radio bursts to wireless systems, we need a way to determine the likelihood of occurrence of bursts exceeding some threshold for harmful effects. An obvious way to do this is to consult the historical record, which was the approach taken by Nita et al. (2002, 2004). However, Figure 5, adapted from Nita et al. (2002), shows that the largest bursts in the 1-2 GHz frequency range appear to be missing from the data. There is a clear cutoff at about $10^5$ sfu in the distribution of bursts occurring at solar maximum. The occurrence of the 2006 December events shows that large bursts do occur, and that the cutoff may be due to instrumental saturation. Even the extrapolated distribution, however, would predict the 2006 December 6 event to occur only once in 100 years at solar minimum. We briefly examine the way these bursts affected the RSTN and OVSA radio instruments, and by implication, how saturation may have affected the historical record.

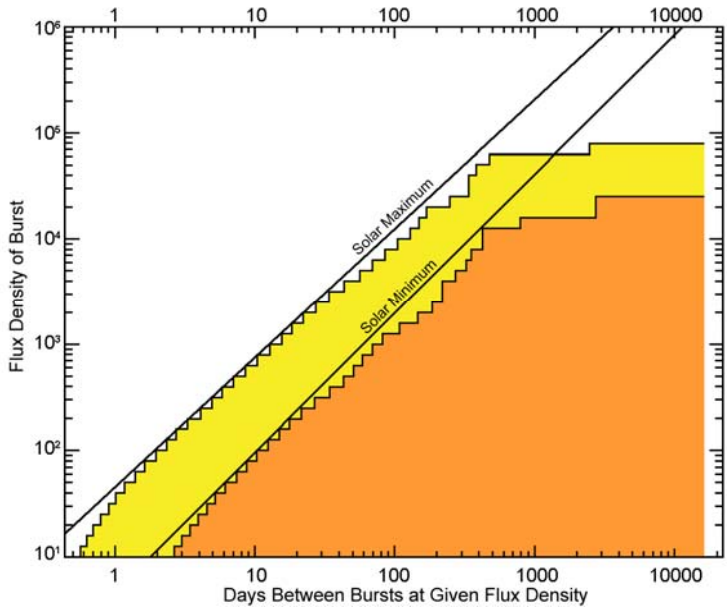

**Figure 5:** Distribution of sizes (peak flux densities) of solar radio bursts over 40 years of burst reports from NOAA, separated into bursts occurring within 2 years of solar maximum and those occurring within 2 years of solar minimum (see Nita et al. 2002 for details). The horizontal scale gives the number of days between bursts. Both distributions show an apparent cutoff in flux density, i.e. the highest flux density bursts are missing. Note that extrapolating the distributions to 106 sfu predicts such bursts should occur once every 10 solar maximum years (about once every 25 years), or once in 41 solar minimum years (about 100 years).

Different receiver designs will be affected by saturation in different ways. The RSTN data from Learmonth for Flare 3 (Fig. 4*d*) shows one way that receivers may react, by simply clipping the signal at their design limit. However, the fluxes reported by RSTN sites for Flares 2 and 4 (Table 1) are far below those measured by OVSA, suggesting a different failure mode for those sites. Unfortunately, the detailed records for these two flares are not yet available at NOAA, so we have only the peak flux reports.

OVSA, too, suffered saturation of at least two different kinds during the largest burst (Flare 2) on 2006 December 6. One type of saturation results in data being unrecoverably lost due to incorrect attenuation inserted into the system. This results in overdriving the A/D converter, and causes the unrecoverable loss of data shown in Fig. 2, especially at 1.2 GHz. The other type of saturation is more subtle, where there is an in-range measurement, but due to non-linear amplifier response the measured flux density is lower than its true flux density. The OVSA measurements are made in parallel on several nominally identical antennas, and by plotting the flux density measured on one antenna versus another we do see such nonlinearity. However, lacking a clear preference for one antenna over another, we use the average value of these independent measurements. Thus, although it could be argued that the antenna measuring the highest flux density suffered the least saturation, we take the more conservative average keeping in mind that the true flux density could be higher.

We now wish to transfer OVSA's absolute flux density scale to the FST data, which are available at much higher time and frequency resolution. To do this, we have to identify the specific times in the FST data and integrate over the same bandwidth as covered by the OVSA receivers. Figure 6 shows a 30 s segment of the FST data at full resolution, with small boxes outlining the durations and bandwidths of individual OVSA measurements at 1.2 and 1.4 GHz. Integrating over these boxes, we obtain the exact data for comparison with OVSA and transfer of the calibration. Figure 7 shows the FST measurements integrated over the regions like the examples in Fig. 6, overplotted with appropriate scaling to match the OVSA data.

Fig. 7 shows that the two instruments agree to a high degree of precision except for the period around 18:50 UT, when the FST gain control malfunctioned for a time. FST uses signals from three of the

- 5 -

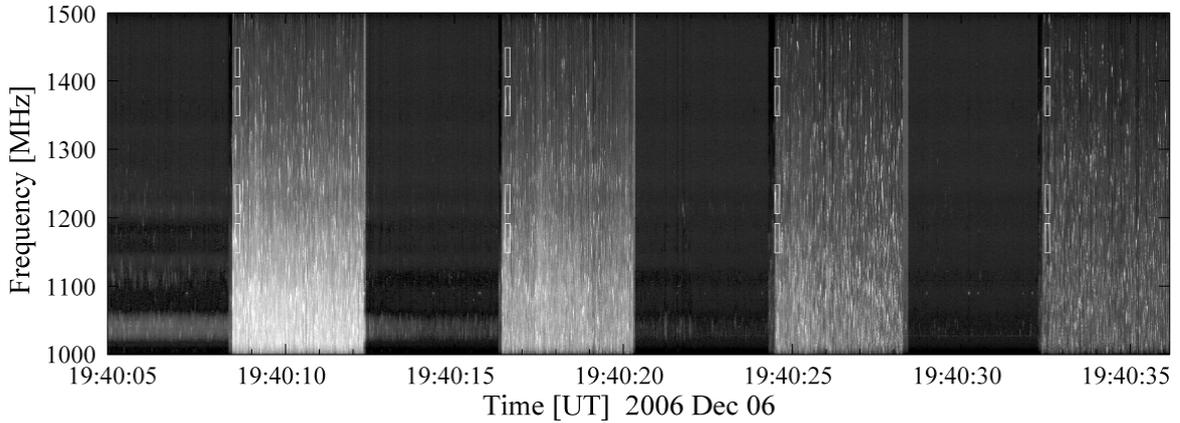

**Figure 6:** A 30 s of period of FST data at full resolution, for Flare 2, showing the times and bandwidths of the OVSA measurements at 1.2 and 1.4 GHz. OVSA is a double-sideband system, so it covers sidebands on either side of the nominal frequency. The alternating brighter and darker bands of 4 s duration (8.1 s period) are due to switching circular polarization. The brighter bands are RCP, and show thousands of densely packed spike bursts. The darker bands are LCP, and are dominated by instrumental artifacts caused by cross-talk from the RCP feed, hence should be ignored. The integrated flux density in such boxes for the entire duration of the burst is shown in Fig. 7.

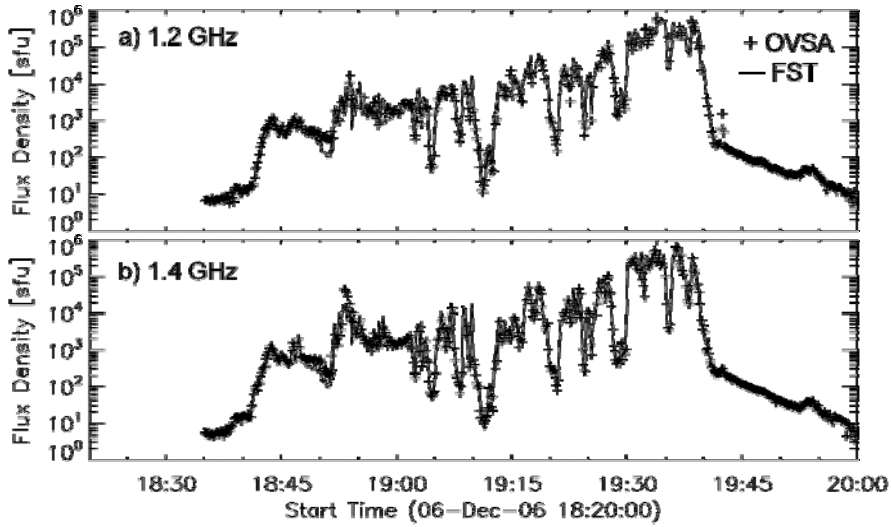

**Figure 7:** Radio flux density at two frequencies for the event of 2006 December 6, comparing the OVSA measurements (+ symbols) with the FST measurements integrated over the identical times and bandwidths as in Fig. 6. The FST data of uncalibrated flux density are scaled with a fixed multiplicative factor to match the OVSA points, which yields the factor needed to transfer OVSA's calibration to FST.

OVSA dishes, but two of the FST receivers suffered saturation during the highest peaks, especially during the period that most affected the GPS receivers, from 19:30-19:40 UT. This causes a visible suppression (not shown) of the peaks in the data from those two antennas relative to the third. The overall peak flux densities for the 2006 December 6 event, as measured with OVSA are $6.5 \times 10^5$ sfu at 1.2 GHz, $1 \times 10^6$ sfu at 1.4 GHz, and $5 \times 10^5$ sfu at 1.6 GHz. However, recall that sfu is a flux density, with units W m$^{-2}$ Hz$^{-1}$, and so these values represent an average flux over the OVSA bandwidth (roughly 100 MHz). The relevant GPS bandwidth is much smaller (1 or 10 MHz, depending on the GPS receiver operation), and as Fig. 6 shows the radio emission occurs in very narrow band, short-duration spikes.

Using FST data, we plot flux densities integrated over 10 MHz in Figure 8 during the highest-flux period, 19:30-19:40 UT. We find that the flux density momentarily reaches peak levels of $1.85 \times 10^6$ sfu at 1.227 MHz (the GPS L2 frequency) and $2.5 \times 10^6$ sfu at 1.405 MHz (for comparison with RSTN). Note that the fluctuations in Fig. 8 are *not* statistical noise. As discussed by Nita et al. (2007), the statistical signal to noise ratio for a single power measurement is exactly 1 for Gaussian noise, so after averaging over $M$ time samples and $N$ frequencies, the signal to noise ratio will be $1/\sqrt{MN}$. Each point in Fig. 8 represents an average over $M = 122$ time samples and $N = 10$ frequencies, for a signal to noise ratio of about 0.03. This is about 20 times smaller than the fluctuations in Fig. 8. The extreme variability of the solar signal on such short time and narrow frequency scales means that gauging the effect of solar radio noise on GPS receivers requires solar observations using high resolution spectrographic data like that provided by FST and other spectrographs around the world.

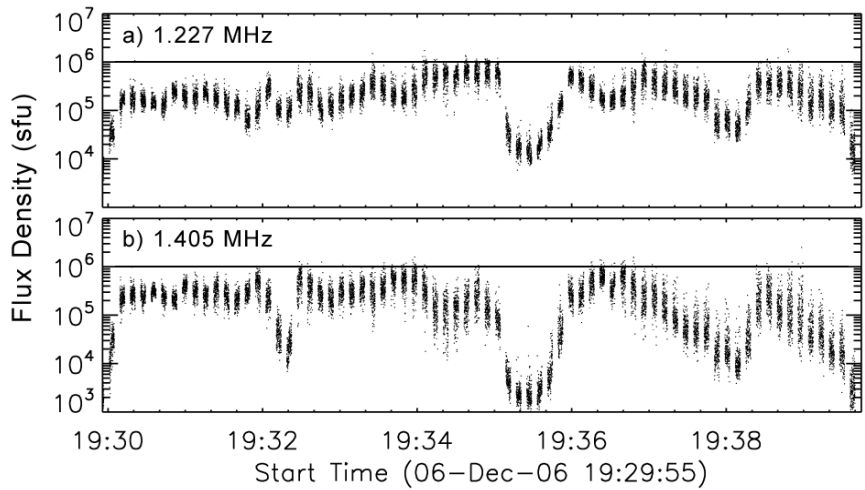

**Figure 8:** Full resolution RCP flux density, integrated over 10 MHz bandwidth *a)* for GPS L2 frequency, *b)* for 1.4 GHz. The gaps are periods when LCP was being measured The variations in each 4 s of data points is due to true variations in the solar signal, and are not statistical noise. Therefore, the points above the $10^6$ sfu level (indicated by the horizontal line in each plot) are significant.

## 4. Emission Mechanism and Physical Conditions Responsible for the Solar Flux Density

From the spectral signature of the emission seen in Fig. 6, especially near the end of the 30 s period shown, we see that the emission is made up of extremely narrow-band (3-5 MHz) spikes of short duration (< 20 ms). Such bursts are well known, and are not surprisingly called spike bursts (Slottje 1978). However, in those bursts where such spikes are seen, generally the number of spikes is in the hundreds to thousands (e.g. Csillaghy & Benz 1993), and although they can occur in clusters, they usually last for only a few minutes and are sufficiently separated that individual spikes can be seen. During more than 40 minutes of the 2006 December 6 event, the spikes were so numerous as to blend into a single highly fluctuating level of emission, only becoming individually distinguishable near the end of the burst, as shown in Fig. 6. A simple estimate of number of spikes (assuming no overlap) to fill a 500 MHz bandwidth for 40 minutes requires more than 10 million spikes to have been produced during this burst.

As mentioned in §1, the ECM mechanism is thought to be responsible spike bursts (see Treumann 2006 for a general review, and Fleishman 2006 for a review of the solar burst case). This coherent mechanism is naturally highly polarized, and is due to high-energy electrons mirroring in the converging field of a closed magnetic loop, where they form a loss-cone distribution that is unstable to catastrophic wave growth (Wu & Lee 1979; Holman et al. 1980; Melrose & Dulk 1982). The emission occurs near the cyclotron frequency or its low harmonics, which requires a magnetic field strength of order 450 G. To maintain the instability requires replenishment of the high-energy electrons, which implies sustained acceleration and excellent trapping over 40 minutes (see Rozhansky et al. 2008, where the local source model is discussed in some detail). Such trapping in turn requires relatively low density in the loop (Fleishman et al. 2003) to avoid collisional energy loss. One can estimate the density assuming that collisional energy loss of electrons of energy ~100 keV (e.g. eq. 4 of Bastian et al. 2007) causes the long exponential decay (time constant about 208 s) visible in Fig. 7 after the spikes turn off around 1940 UT. The implied density is around $10^9$ cm$^{-3}$, and even lower for lower energy electrons; hence such loops are generally not visible in EUV or X-ray images. The instability in a given volume of plasma is quickly quenched (within < 20 ms, which accounts for the short duration of the spikes), so to create 10 million spikes requires that these appropriate conditions be satisfied again and again, at many places in the loop simultaneously. One might consider such conditions to occur only very rarely, except that we know that they occurred three times within one week in AR 10930. This strongly suggests that AR 10930 maintained a magnetic configuration highly favorable for the production of ECM emission. Because the historical record appears to be incomplete, it remains uncertain how often such conditions occur.

## 5. Conclusion

We have given an overview of the flares in AR 10390 during 2006 December, have placed these bursts in the context of the historical record, and have attempted to obtain a reliable L-band flux density for the largest of the events, the event of 2006 December 6. We have also briefly examined the solar conditions causing the extreme solar radio flux densities observed. Our main conclusions are as follows:
- The cause of the extreme flux density is an unusually high density of spike bursts due to the Electron-Cyclotron Maser mechanism.

- The December 6 event reached a flux density of at least $10^6$ sfu at 1.4 GHz, momentarily reaching flux densities as high as $2.5\times10^6$ sfu during individual spikes.
- Bursts of $10^6$ sfu, *if they follow the general size distribution of solar bursts*, should occur once every 25 years at solar maximum rates, or once every 100 years at solar minimum rates. The events of 2006 December 6 are therefore extremely rare occurrences.
- In the historical record, however, the largest bursts are missing, and we argue that it may be due to instrumental saturation. This leaves open the question of how often such large events may occur.
- To monitor the effects of solar bursts on GPS, GLONASS, and Galileo navigation systems, a world-wide network of solar spectrographs is needed that can measure right-circularly polarized emission at L-band with high frequency and time resolution, without saturation.

As our society becomes ever more dependent on wireless technology, the effects of solar radio bursts can be expected to appear more often. Mission-critical systems should be designed with solar radio emission in mind. A warning system based on an improved set of world-wide instrumentation could be implemented at relatively low cost, taking advantage of new technology that allows broadband digital signal measurements. Ultimately, such extreme flux density bursts need to be studied at high spatial resolution with arrays such as the proposed Frequency Agile Solar Radiotelescope (FASR) in order to better understand the conditions leading to their occurrence, and ultimately to be able to predict such events.

## 6. Acknowledgments

This work was supported by NSF grants ATM-0707319, AST-0352915 and AST-0607544, and NASA grant NNG06GJ40G to New Jersey Institute of Technology.